\documentclass[journal]{IEEEtran}
\def\comment#1{}
\usepackage{graphicx,amsfonts,amsmath,amssymb,xcolor,amsthm}
\usepackage{algorithm,algorithmic,bm,color}
\usepackage{multirow}
\usepackage{hyperref}
\usepackage[noadjust]{cite}
\usepackage{adjustbox}
\usepackage{makecell}
\usepackage{stackengine}
\newcommand\xrowht[2][0]{\addstackgap[.5\dimexpr#2\relax]{\vphantom{#1}}}
\def\comment#1{}
\newcommand{\argmin}{\arg\!\min}

\newcommand{\Emat}[0]{{{\bf E}}}

\newcommand{\Mmat}[0]{{{\bf M}}}

\newcommand{\Rmat}[0]{{{\bf R}}}

\newcommand{\Xmat}{{\bf X}}
\newcommand{\Ymat}[0]{{{\bf Y}}}

\newcommand{\ev}[0]{{\boldsymbol{e}}}

\newcommand{\mv}[0]{{\boldsymbol{m}}}

\newcommand{\vv}{\boldsymbol{v}}

\newcommand{\xv}{\boldsymbol{x}}
\newcommand{\yv}{\boldsymbol{y}}

\newcommand{\Phimat}{\boldsymbol{\Phi}}

\newcommand{\ts}{^{\top}}
\newcommand{\inv}{^{-1}}
\newcommand{\ie}{{\em i.e.}}
\newcommand{\eg}{{\em e.g.}}
\newcommand{\etc}{{\em etc}}

\begin{document}
\title{Block Modulating Video Compression:\\ An Ultra Low Complexity Image Compression Encoder for Resource Limited Platforms}

\author{
	\IEEEauthorblockN{Siming Zheng, Yujia Xue, Waleed Tahir, Zhengjue Wang, Hao Zhang, Ziyi Meng, Gang Qu, Siwei Ma and \\ Xin Yuan, {\em Senior Member, IEEE}} 
\thanks{ 
\newline Yujia Xue and Waleed Tahir are with Department of Electrical and Computer Engineering, Boston University, Boston, MA 02215, USA.
\newline Siming Zheng is with Computer Network Information Center, Chinese Academy of Sciences, Beijing 100190, China and also with University of Chinese Academy of Sciences, Beijing 100049, China.
\newline Zhengjue Wang is with Department of ECE, New Jersey Institute of Technology, Newark, New Jersey 07102, USA.
\newline Hao Zhang is with Department of Population Health Sciences. Weill Cornell Medicine, New York, NY 1006, USA. 
\newline Ziyi Meng is with State Key Laboratory of Information Photonics and Optical Communications, Beijing University of Posts and Telecommunications,
Beijing 100876, China. 
\newline{Xin Yuan (corresponding author) was with Bell Labs, 600 Mountain Avenue, Murray Hill, NJ 07974, USA.
He is now with Westlake University, Hangzhou, Zhejiang 310024, China. Email: xyuan@westlake.edu.cn.} 
\newline Yujia Xue and Waleed Tahir conducted related work when they were summer interns at Bell Labs in 2021 and 2020, respectively.}
}	

\markboth{Journal of \LaTeX\ Class Files,~Vol.~XX, No.~XX, October~2021}%
{Shell \MakeLowercase{\textit{et al.}}: Bare Demo of IEEEtran.cls for IEEE Journals}
%



\maketitle
\begin{abstract}
We consider the image and video compression on resource limited platforms.
An ultra low-cost image encoder, named Block Modulating Video Compression (BMVC) with an encoding complexity ${\cal O}(1)$ is proposed to be implemented on mobile platforms with low consumption of power and computation resources.
We also develop two types of BMVC decoders, implemented by deep neural networks.
The first BMVC decoder is based on the Plug-and-Play (PnP) algorithm, which is flexible to different compression ratios. 
And the second decoder is a memory efficient end-to-end convolutional neural network, which aims for real-time decoding.
Extensive results on the high definition images and videos demonstrate the superior performance of the proposed codec and the robustness against bit quantization.
\end{abstract}


%
\IEEEpeerreviewmaketitle

\section{Introduction \label{Sec:intro}}
\IEEEPARstart{I}{mage} and video compression algorithms have been developed for about 30 years while state-of-the-art codecs are still mainly based on the MPEG (Moving Picture Experts Group) structure~\cite{LeGall_91_MPEG}, which was originally developed for broadcasting.
However, 30 years ago, videos were usually captured and produced in studios and thus the encoding time and cost can be very long and expensive;
while the decoder at the customer's end needs to be of low complexity since it is in every family (on the television).
Nowadays, image and video codecs are all over mobile platforms like cellphones, drones, \etc., where images and videos can be captured anywhere, anytime, given enough power.
Moreover, the devices (\eg, cellphones, laptops, desktops) we now use to decode image and video streams have by magnitude higher computation power than decades ago.
Therefore, now it is natural for the image and video compression to evolve towards the combination of a low-cost encoder and a (possibly) computationally heavy decoder, where the low-cost encoder fits better on those {\em resource limited robotic platforms}.

This paper considers the image and video compression codec on those mobile platforms with sensitive constraints on battery, computation, and bandwidth.
Specially, we highlight drones and robotics as representative applications.
In these use cases, {\em only the low-cost encoder needs to be implemented in real-time on the mobile platform, but the decoding can happen after transmission on other tabletop platforms} such as edges~\cite{Shi_16Edge} or cloud.
Since most of these mobile platforms are running on standalone batteries, a power saving on the encoder can extend the running time of other sensors and motion modules on drones or robotics, which is of significant interest in extreme cases such as moon rovers and other military applications.

Bearing this concern in mind, we propose an ultra low-complexity image and video codec using block modulation, dubbed BMVC representing Block Modulating Video Compression codec. 
The underlying principle of BMVC is to mask the high resolution image (via binary random coding patterns composed of \{0,1\}) and then decompose it into different small (modulated) blocks.
Finally, these blocks are summed up to {\em a single block} and quantized as the compressed signal to be transmitted. 
Since no multiplication is involved during this encoding process, the complexity of the BMVC encoder is way lower than MPEG-based codecs.
Moreover, the summation over modulated image blocks by binary masks can be essentially implemented as summations according to a pre-defined look-up table during this encoding process.
Therefore, the complexity of the BMVC encoder can be ${\cal O}(1)$.
Hereby, the mask pattern plays the role of basis or key during the compression.
Without loss of generality, we use random binary patterns with equal probability being 1 or 0, which is stored as a look-up table on the mobile platform.

\begin{figure*}
\centering
\includegraphics [width=2.0\columnwidth]{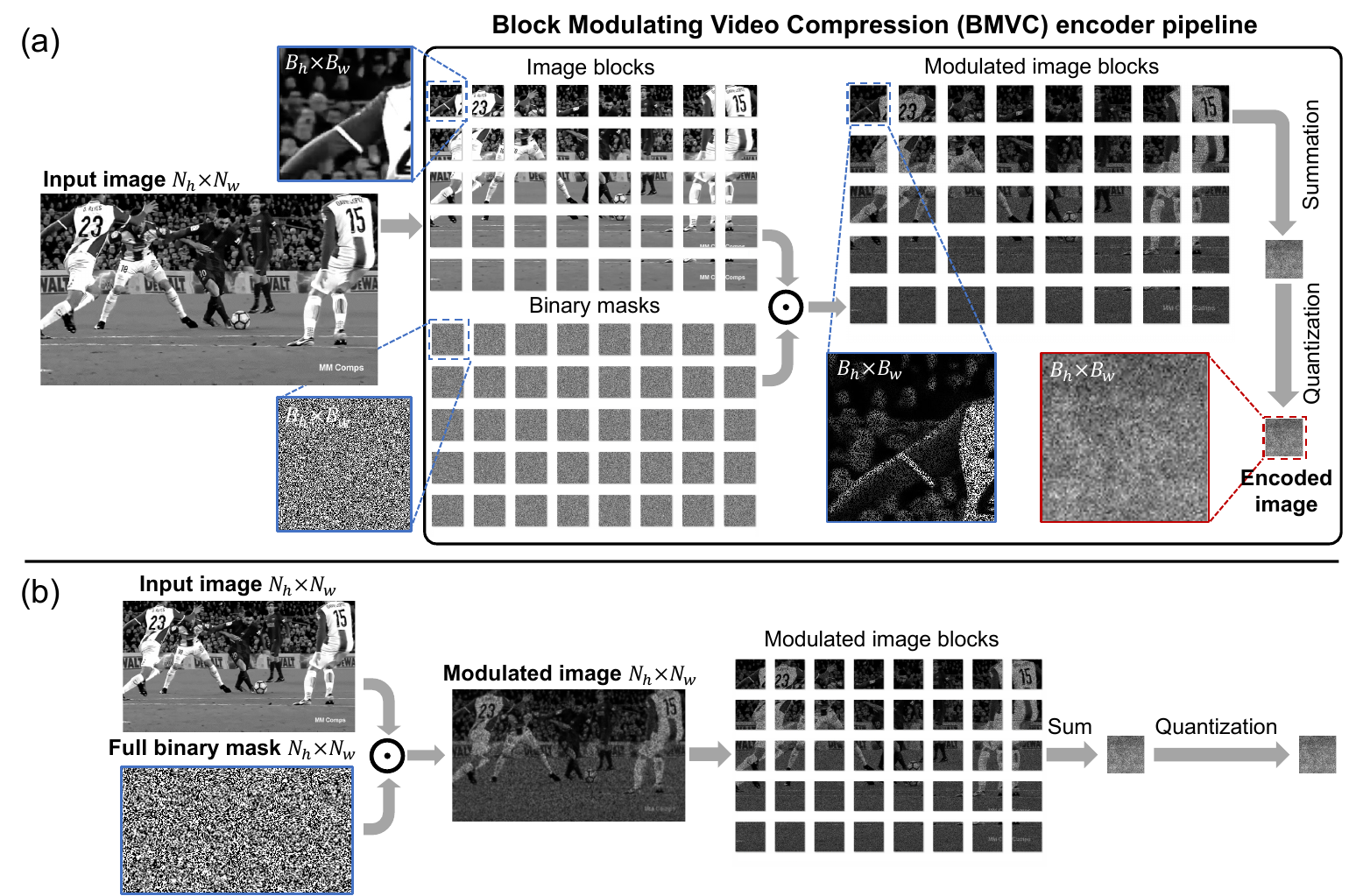}
\caption{Pipeline of the proposed Block Modulating Video Compression (BMVC) encoder. (a) For each input image with size of $N_h \times N_w$. The input frame is first divided into non-overlapping blocks of size $B_h\times B_w$. Then the image blocks are modulated (element-wise multiplication) by binary masks of the same size. Next, these modulated image blocks are summed together to yield a single block of size $B_h\times B_w$. At last, the summed modulated image block is quantized to a user-defined bit depth ($8\sim 16$-bit) and then transmitted to the receiver. An equivalent encoding pipeline is shown in (b) where the modulation happens before dividing into blocks.}
\label{fig:BMVC_encoder}
\end{figure*}

\subsection{BMVC Pipeline: Key Idea}
Figure~\ref{fig:BMVC_encoder} shows the basic principle of the proposed BMVC encoder, where the input is a raw image captured by the sensor, \eg, a Charge-Coupled Device (CCD) or Complementary Metal–Oxide–Semiconductor (CMOS) camera, with a spatial resolution of $N_h\times N_w$ pixels, where $N_h$ denotes the height and $N_w$ denotes the width, respectively\footnote{Hereby, we consider the grayscale images. BMVC can be extended to color images on the Bayer patterns or imposed on Red, Green and Blue channels separately or on the Y, U and V channels, respectively.}. 
A binary random mask of the same image size composed of \{0,1\} with each one of half probability, is stored on the mobile platform and plays the role of a key (or called basis) in BMVC.
Both the HD image and the mask are divided into {\em non-overlapping}\footnote{Overlapping blocks can also be used in our BMVC framework with minimum changes in the decoder and the performance is similar to the non-overlapping case.} blocks of size $B_h\times B_w$.
Following this, the mask blocks are superimposed onto each image block to yield the {\em modulated image blocks} with the same block size but half number of pixels are missing.
Next, these modulated blocks are summed into a single block, which is then quantized to the encoded image block.
This quantized single block is the only information we need to transmit after encoding, \ie, the output of the encoder. Note that the masks are pre-defined and shared between the encoder and decoder, and thus do not need to be transmitted.

Hereby, the number of divided image blocks defines the compression ratio (Cr) of this encoding process
 ${\rm Cr}=\frac{N_h N_w}{B_h B_w}$.
Notably, the summation and quantization may hurt the dynamic range of the compressed data.
While we later show the BMVC pipeline is {\bf robust against different quantization bits} and able to provide consistent performance in a wide range of compression ratios.

On the decoder side, after receiving the encoded block, the BMVC decoding algorithms are employed to reconstruct the original HD image, provided the masks {\em known a priori} used in the encoder.
In this paper, we consider two reconstruction algorithms, one being the Plug-and-Play (PnP)~\cite{Chan2017PlugandPlayAF,Ryu2019PlugandPlayMP,Venkatakrishnan_13PnP,Sreehari16PnP} optimization algorithm (BMVC-PnP) with a deep denoising neural network in each iteration~\cite{Yuan2020_CVPR_PnP} and the other being an end-to-end convolutional neural network (BMVC-E2E) for real-time video decoding.

\subsection{Contributions}
We want to emphasize that BMVC is not designed to replace the current codec standard, but to provide an alternative option for platforms under extreme power/computation constraints.
Here we summarize the contributions of this work as follows:
\begin{itemize}
    \item A brand new image and video compression codec is proposed for resource limited platforms.
    \item The proposed BMVC encoder has extremely low computation cost.
    \item Two high quality BMVC decoders are developed with each enjoying different benefits, \ie, the BMVC-PnP features flexibility and BMVC-E2E for real-time applications.
    \item The proposed BMVC codec is fully evaluated by different compression ratios, different quantization bits and benchmarked with other compression methods.
    \item Extensive results demonstrate the superior and consistent performance of BMVC on resource limited platforms. 
\end{itemize}

\subsection{Organizations of the Paper}
The remaining part of this paper is organized as follows. Section~\ref{Sec:background} reviews the related background knowledge that BMVC builds on. Section~\ref{Sec:BMVC_model} describes the mathematical model of the encoding process of BMVC and also develops two high quality decoders.  Extensive results are provided in Section~\ref{Sec:Results} and Section~\ref{Sec:Conclusion} concludes the entire paper and discusses the applications. 

\section{Background Knowledge \label{Sec:background}}
Similar to other codecs, the proposed BMVC pipeline is composed of an encoder and a decoder. As shown in Fig.~\ref{fig:BMVC_encoder}(a), the encoder consists of blocking, masking, summation and quantization.
It is worth noting that the blocking and masking steps can be switched; specifically, a mask of size $N_h\times N_w$ can be used to first mask the entire image (or an HD frame) and then the modulated image (of size $N_h\times N_w$) can be divided into blocks of size $B_h \times B_w$; following this, these modulated blocks are summed to a single block.
This alternative interpretation of BMVC gives us the benefit that only one full-frame mask of size $N_h\times N_w$ is needed to be pre-defined and stored before deployment, shown in Fig.~\ref{fig:BMVC_encoder} (b). 

One underlying principle of the BMVC encoder is compressive sensing (CS)~\cite{Candes06_Robust,Donoho06_CS}, where a small number of measurements can be used to reconstruct the full signal with a higher dimension under some mild conditions.
Specifically, BMVC shares the same spirit with snapshot compressive imaging (SCI)~\cite{Yuan2021_SPM}.
In the following, we review the basic idea of SCI and deep learning methods for CS inversion, \ie, the reconstruction process. 

\subsection{Snapshot Compressive Imaging}
Snapshot compressive imaging utilizes a 2D detector to capture high-dimensional ($\ge3$D) data.
It was first developed to capture high-speed videos (spatio-temporal datacubes) with a relatively low-speed camera~\cite{Llull13_OE_CACTI,Yuan14CVPR,Qiao2020_APLP} or to capture hyperspectral images (spatio-spectral datacubes) in a single-shot~\cite{Wagadarikar08_CASSI,Meng20ECCV_TSAnet} among other applications~\cite{Qiao2020_CACTI,Llull15Optica,Tsai15OL,Sun16OE,Qiao2021_TCI_CSOCT,Sun17OE}.
Generally, the fundamental idea of SCI is to modulate each low dimensional slice (frames or spectral channels) from an underlying high-dimensional data structure with a different mask and then sum these modulated slices into a single low dimensional measurement. 

In our proposed BMVC encoding process, as shown in Fig.~\ref{fig:BMVC_encoder}, we can interpret its forward model as a slightly different way following the same spirit of SCI.
Consider the image blocks in BMVC as low dimensional slices in the SCI modality, and the modulation masks corresponding to the SCI masking scheme.
Then the BMVC modulated blocks are summed to a single block measurement, which corresponds to the compressed measurement in SCI.
From this perspective, the encoding process of BMVC is essentially the same as SCI. 
Whereas, there is a key difference between BMVC and SCI.
That is in SCI, the frames in the video sequence or the spectral channels in the hyperspectral datacube are strongly correlated as they share highly similar spatial features.
While in BMVC, each block corresponds to {\em a different portion of the image and these blocks are not necessarily correlated}, which makes the BMVC decoding more challenging than SCI modalities.
Though the decoding task poses a big challenge, in this paper, we show that using advanced deep-learning-based CS inversion algorithms, high resolution images can still be faithfully decoded from the highly ill-posed, low-cost BMVC encoder.

\subsection{Deep Learning for SCI Inversion}
The inverse problem of compressive sensing is ill-posed, as there are more unknown parameters to be estimated than known measurements.
Towards this end, different priors have been employed as regularizations to help solve the CS inverse problem.
Widely used priors include sparsity~\cite{Duarte08SPM} and piece-wise smoothness such as total variation (TV) and low rankness of similar image patches.
Implicit regularizations have also been explored by using standard denoising algorithms (such as NLM~\cite{buades2005non}, BM3D~\cite{dabov2006image}) as plug-and-play priors~\cite{lai2020single}. 
Other algorithms have also been used to solve the SCI reconstruction, such as TwIST~\cite{Bioucas-Dias2007TwIST}, GAP-TV~\cite{Yuan16ICIP_GAP} and DeSCI~\cite{Liu19_PAMI_DeSCI}.

Recently, deep learning (DL) has been used for solving inverse problems in computational imaging systems for high quality and high-speed reconstruction.
Specially, existing DL-based inversion algorithms can be categorized into three different classes~\cite{Yuan2021_SPM}: $i$) end-to-end convolutional neural networks (E2E-CNN) for high-speed reconstruction~\cite{Qiao2020_APLP,Meng20ECCV_TSAnet,Miao19ICCV,Cheng2021_CVPR_ReverSCI,xue2019reliable,li2018deep}, $ii$) deep unfolding/unrolling networks~\cite{Ma19ICCV,Wang19_CVPR_HSSP,Li2020ICCP,Huang2021_CVPR_GSMSCI,gregor2010learning,Yang_NIP16_ADMM-net,Metzler_17NIPS_LDAMP} with interpretability, and $iii$) the Plug-and-Play (PnP) algorithms~\cite{Yuan2020_CVPR_PnP,Zheng20_PRJ_PnP-CASSI,PnP_SCI_arxiv2021} using pre-trained denoising networks as implicit priors to regularize inverse problems.  

For our BMVC decoding case considered in this work, due to the large scale of the data, we employ two decoding algorithms: a plug-and-play optimization-based decoder (BMVC-PnP) and an end-to-end feed-forward decoder (BMVC-E2E).
Regarding the data dimension, we consider the HD data of size\footnote{We use $1080\times 1920$ as an example to demonstrate the idea of BMVC due to its wide usage. BMVC is ready to be used to other size of images and videos. In fact, there is no constraints on the input size of BMVC codec.} $1080\times 1920$, which is decomposed into blocks of various sizes with a wide range of Cr from 24 to 150.
Given such a wide Cr range, it is challenging to train a deep unfolding network with multiple CNNs for this large-scale data due to the limited GPU memory.
For applications where real-time reconstruction is not necessary, we first develop an iterative BMVC decoder based on the plug-and-play algorithm.
The BMVC-PnP decoder employs a pre-trained FFDNet~\cite{Zhang18TIP_FFDNet} as the denoiser to regularize the inversion, which makes the PnP framework flexible to different Crs.
On the other hand, there still exist many applications that require real-time decoding.
To achieve a more practically usable BMVC codec, we further develop a non-iterative, feed-forward decoder based on an end-to-end CNN.
It is worth mentioning that even for the end-to-end approach, the training of BMVC-E2E is not trivial since it requires an ultra deep CNN which cannot be fit in the memory of a single GPU.
Fortunately, recent work on reversible networks~\cite{Kellman20_ResverNet} has shown to reduce the memory requirement during the training phase and thus we build our BMVC-E2E decoder around 3D reversible convolution modules~\cite{Cheng2021_CVPR_ReverSCI}. 
Finally, the BMVC-E2E decoder is capable of decoding images at a frame rate of $4\sim 14$ frames per second ($fps$)  depending on the Cr with a single decent GPU. 

\section{Block Modulating Video Compression \label{Sec:BMVC_model}}
In this section, we elaborate the mathematical details of the BMVC pipeline. In particular, we use grayscale images as an example to derive the mathematical model. As mentioned before, BMVC is ready to handle RGB color images and videos, where we can conduct BMVC on all RGB channels or YUV (or YCbCr) channels. In our experiments, we have further found that decent decoding results can be obtained by only performing BMVC on the Y channel and a simple downsampling/upsampling can be used for the U and V channels. 

Let $\Xmat \in {\mathbb R}^{N_h\times N_w}$ denote the input HD image (left-middle in Fig.~\ref{fig:BMVC_encoder}(a)).
The first step is to divide the input image $\Xmat$ into $N_b$ non-overlapping small blocks of size $\Xmat_b \in {\mathbb R}^{B_h \times B_w}$, where $B_h < N_h $ and $B_w < N_w$.
It is straightforward to show that the number of blocks equals to the compression ratio $N_b$ = Cr.
Then the image blocks are element-wise modulated by the binary masks, which gives us
\begin{equation}
\tilde{\Xmat}_b = \Xmat_b \odot \Mmat_b, \quad \forall b = 1,\dots, N_b,
\label{eq:element_p}
\end{equation}
where $\odot$ denotes the element-wise multiplication and $\Mmat_b \in \{0,1\}^{B_h\times B_w}$ denote the binary masks shown in the middle-bottom in Fig.~\ref{fig:BMVC_encoder}(a).

Note that in practice, Eq.~\eqref{eq:element_p} can be performed by summations according to a look-up table rather than any multiplication.
As a result, the actual computational complexity of this BMVC encoder is as low as ${\cal O}(1)$.
Detailed analysis on the BMVC complexity can be found in subsection~\ref{subsec-complexity}.

Generally, the number of blocks $N_b$ can be obtained by 
\begin{equation}
N_b = \left\lceil \frac{N_h}{B_h}\right \rceil \left \lceil \frac{N_w}{B_w} \right\rceil,
\end{equation}
where $\lceil a \rceil $ denotes the minimum integer no smaller than $a$.
For optimal BMVC encoding and decoding performance, we choose both $B_h$ and $B_w$ to be divisors of $N_h$ and $N_w$ so that we can maximize compression ratio by having non-overlapping blocks.
Note that we only need to save the indices of each block, and then to stitch them together to get the full-sized image back in the decoder.

\subsection{Compressed Measurement}
After blocking and binary modulation, the next step is to sum all these modulated blocks to yield a single compressed measurement, \ie,
\begin{equation}
\Ymat_0 = \sum_{b=1}^{N_b} \tilde{\Xmat}_b = \sum_{b=1}^{N_b} \Xmat_b \odot \Mmat_b.  \label{eq:encoder_sum}
\end{equation}
This $\Ymat_0 \in {\mathbb R}^{B_h \times B_w}$ is the {\em noise-free compressed measurement} that we are going to transmit. 

Before sending the measurement, the last step is bit quantization, which imposes additional quantization error to the actual compressed signal $\Ymat$.
In this case, we have the forward model of 
\begin{equation}
   \Ymat =  \sum_{b=1}^{N_b} \Xmat_b \odot \Mmat_b + \Emat_q, 
   \label{Eq:YXM_E}
\end{equation}
where $\Emat_q\in {\mathbb R}^{B_h \times B_w}$ denotes the quantization error.
In our experiments, we performed $8\sim 16~bit$ quantization and found the BMVC pipeline is robust against quantization error and provides consistent reconstruction quality.

\subsection{Forward Model in Vectorized Formulation}
For the sake of describing the reconstruction algorithm employed in the decoder, we hereby introduce the vectorized formulation of \eqref{eq:encoder_sum}.
Let
\begin{eqnarray}
\xv_b &=& {\rm Vec} (\Xmat_b), \\
\mv_b &=& {\rm Vec} (\Mmat_b),\\
\yv &=& {\rm Vec} (\Ymat),
\end{eqnarray}
where the ${\rm Vec}(~)$ operator vectorizes the ensued matrix by stacking its columns. 

After vectorization, Eq.~\eqref{eq:encoder_sum} can be re-formulated as
\begin{equation}
\yv_0 = \underbrace{\left[{\rm Diag}(\mv_1), {\rm Diag}(\mv_2),\dots, {\rm Diag}(\mv_{N_b})\right]}_{\stackrel{\rm def}{=} \Phimat \in {\mathbb R}^{B_hB_w \times (B_h B_w N_b)}} \underbrace{\left[\begin{array}{c}
\xv_1\\
\xv_2\\
\vdots\\
\xv_{N_b}
\end{array}\right]}_{\stackrel{\rm def}{=} \xv \in {\mathbb R}^{B_h B_w N_b}},
\label{Eq:yPhix}
\end{equation}	
where ${\rm Diag}(~)$ embeds its input vector to the main diagonal of a square diagonal matrix.

Given the structure of the forward model $\Phimat$, note that $\Phimat\Phimat\ts$ is a diagonal matrix as
\begin{eqnarray}
\Rmat &\stackrel{\rm def}{=} &\Phimat\Phimat\ts = {\rm Diag}(r_1, \dots, r_{B_hB_w}),  \label{Eq:Rmat}\\
r_i &=&   \sum_{b=1}^{N_b} m^2_{i,b}, \quad \forall i = 1,\dots, B_hB_w,
\end{eqnarray}
where $m_{i,b}$ is the $i$-th element in $\mv_b$.

When the quantization error $\ev_q$ is considered,  the forward model in~\eqref{Eq:YXM_E} can be written as
\begin{equation}
    \yv = \Phimat\xv + \ev_q. \label{Eq:Y_Phix}
\end{equation}
It is worth noting that~\eqref{Eq:Y_Phix} has the formulation of CS, but with a sensing matrix $\Phimat$ being a concatenation of diagonal matrices. 
The performance bound of this special sensing matrix was analyzed in~\cite{Jalali19TIT_SCI}.

Though there is a matrix multiplication in~\eqref{Eq:Y_Phix}, during the encoding process, there is only masking (implemented by a look-up table) and summation as in~\eqref{eq:encoder_sum}.
Again, the complexity of our BMVC encoder is of ${\cal O}(1)$. 

Note that as mentioned in the introduction, only a single full-sized mask is needed to be pre-defined and stored for both encoder and decoder. This mask can be performed on consequent frames in a video sequence.  
This mask can also be designed for different user applications or designed for encryption purpose, for which only authorized person can access the mask and then decode the video, while the compressed measurement is actually encrypted. This provides another benefit of BMVC.  

\begin{figure}[htbp!]
\centering
\includegraphics [width=1.0\columnwidth]{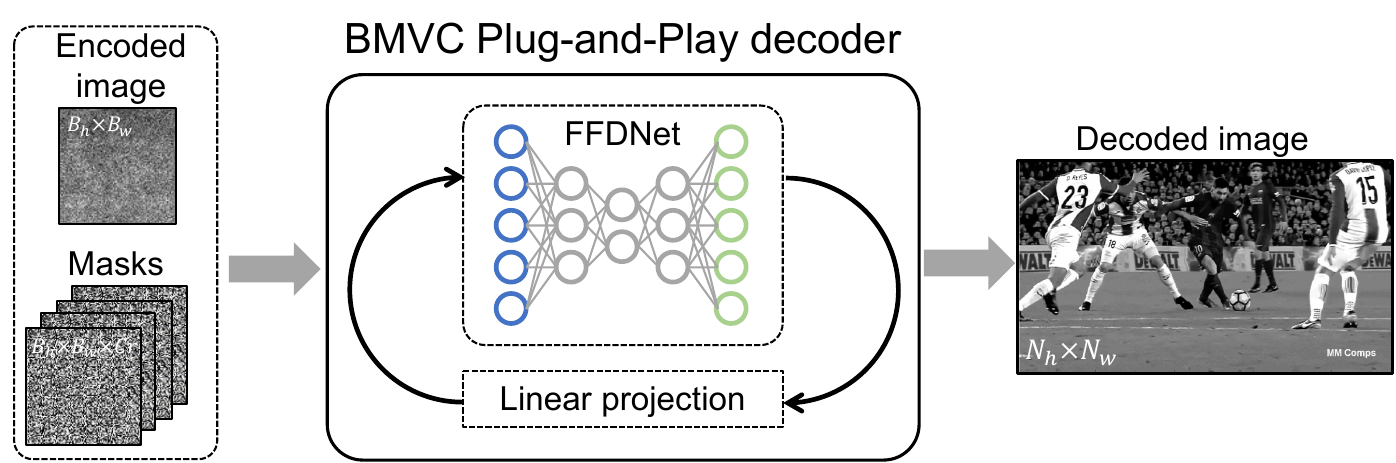}
\caption{Plug-and-Play optimization-based decoding algorithm for Block Modulating Video Compression (BMVC-PnP). The encoded image block along with the modulation binary masks are fed into the BMVC-PnP decoder as inputs. The BMVC-PnP iteratively performs a linear projection step to account for the BMVC encoding process and a deep-learning-based denoising step as an implicit prior. We use a pre-trained FFDNet~\cite{Zhang18TIP_FFDNet} as the denoising CNN for its flexibility and robustness against various noise levels.}
\label{fig:BMVC_PNP}
\end{figure}

\subsection{PnP-based Decoder of BMVC}
So far, we have shown that the BMVC encoder has an extremely low complexity, which makes it perfect for use cases on resource limited mobile platforms.
On the other hand, the decoding process of BMVC becomes highly challenging due to the huge dimensionality mismatch (up to $N_b$ or Cr times) between the encoded block and the original HD image.

Here we first discuss an optimization-based decoder that utilizes the plug-and-play~\cite{Venkatakrishnan_13PnP,Sreehari16PnP} algorithm.
When the decoder receives the encoded image block from the encoder, the goal is to {\em reconstruct} the desired image, provided the modulation masks.
Following the formulation in \eqref{Eq:Y_Phix}, the decoding is an ill-posed problem and thus similar to CS, priors need to be employed in the optimization,
\begin{equation}
    \hat{\xv} = \argmin_{\xv} \|\yv - \Phimat \xv\|_2^2 + \tau R(\xv),
\end{equation}
where $R(\xv)$ denotes the prior term to regularize the inverse problem.
As mentioned in the background knowledge, various priors have been developed for CS and thus can be used for the BMVC decoding.
Recently, the {\em deep prior} learned by neural networks has been used to solve the SCI inversion, leading to the plug-and-play algorithm~\cite{Yuan2020_CVPR_PnP}.

Since the specific structure of the BMVC encoding operator $\Phimat$ (shown in~\eqref{Eq:yPhix}) shares the same mathematical form as SCI modality, the BMVC-PnP decoder has a similar structure as the PnP-GAP~\cite{Yuan2020_CVPR_PnP} framework to solve the inversion problem of BMVC decoding, and here GAP denotes generalized alternating projection~\cite{Liao14GAP,Yuan16ICIP_GAP}.

To be concrete, the BMVC-PnP is an iterative decoder.
Starting from $\vv^{(0)}$, where the superscript denotes the iteration number, the BMVC-PnP is composed of the following two steps:
\begin{eqnarray}
\xv^{(j+1)} &=& \vv^{(j)} + \Phimat\ts(\Phimat\Phimat\ts)\inv (\yv - \Phimat  \vv^{(j)}), \label{Eq:pnpGAP1}\\
\vv^{(j+1)}&=& {\rm Denoise}(\xv^{(j+1)}). \label{Eq:pnpGAP2}
\end{eqnarray}
Due to the diagonal structure of $\Rmat = \Phimat\Phimat\ts$ shown in~\eqref{Eq:Rmat}, \eqref{Eq:pnpGAP1} can be computed efficiently (element-wise operation). 
Regarding \eqref{Eq:pnpGAP2}, the key idea of PnP is that various denoiser can be used to regularize the outcome from the linear step \eqref{Eq:pnpGAP1} to achieve better results. 
Recently, it has been shown that the fast and flexible denoising network, FFDNet~\cite{Zhang18TIP_FFDNet}, is flexible in terms of noise levels and can provide excellent results for different image processing tasks.  
Therefore, in this work we adopt the PnP framework with FFDNet as the BMVC-PnP decoder. 

Though the CNN based FFDNet is very efficient for denoising, the BMVC-PnP decoder is still an iterative algorithm and thus it cannot provide real-time results.
For instance, in the experimental results, we let the BMVC-PnP decoder run for 60 iterations while gradually decrease the $\sigma$ value in the FFDNet step ($\sigma = [20, 10, 5]$, each for 20 iterations).
Finally, the runtime of the BMVC-PnP decoder is about one minute per HD image.

\begin{figure*}
\centering
\includegraphics [width=2.0\columnwidth]{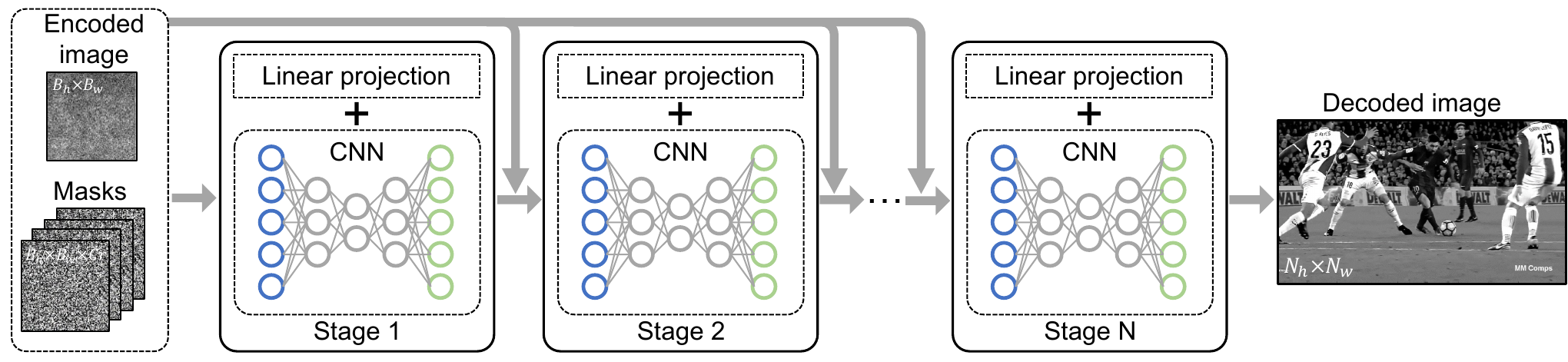}
\caption{End-to-End neural-network-based decoding algorithm for Block Modulating Video Compression (BMVC-E2E). The encoded image block along with the modulation binary masks are fed into the BMVC-E2E decoder as inputs. The feed-forward BMVC-E2E decoder consists of several stages, where each stage contains a linear projection step and a convolutional neural network. All BMVC-E2E decoders are trained in an end-to-end fashion. We use 2D-U-Net and 3D-CNN with reversible blocks (RevSCI) to facilitate memory-efficient training. Detailed network structures can be found in subsection~\ref{Sec:network}.} 
\label{fig:BMVC_E2E}
\end{figure*}

\subsection{End-to-End CNN Decoder of BMVC}
\label{sec-e2e}
To address the speed issue and enable real-time BMVC applications, we present a second BMVC decoder which employs an end-to-end CNN architecture for faster BMVC decoding.
In the following context, we will term the CNN-based end-to-end BMVC decoder as BMVC-E2E.

In order to make the BMVC-E2E robust and interpretable, we design the feed-forward CNN architecture based on unrolling the PnP optimization framework to a few stages~\cite{hershey2014deep_unfold}, as shown in Fig.~\ref{fig:BMVC_E2E}.
Similarly, each stage now contains a linear projection operator to account for the BMVC forward encoding process, and a CNN to serve as an implicit regularization.
Note that though the BMVC-E2E follows the structure of an unrolled optimization, it has two unique features making it different from the PnP approach.
First, unlike the CNN in the PnP approach which is independently trained as an {\em ad hoc denoiser}, the BMVC-E2E decoder is trained in an end-to-end (multiple stage jointly) fashion to perform direct decoding.
Second, remind that the whole purpose of the BMVC-E2E decoder is to accelerate the inference time towards real-time decoding.
The BMVC-E2E needs only a few stages (usually $2\sim 3$ stages are already enough) while the BMVC-PnP generally requires $\ge 60$ iterations to converge.
These unique features of the end-to-end decoder make real-time BMVC decoding possible on standard GPU servers.

The linear projection step follows the identical derivation as in~\eqref{Eq:pnpGAP1}.
While the general structure of deep unfolding is easy to understand, how to design the network in each stage is extremely challenging for efficient BMVC decoding.
As mentioned before, different from other applications of SCI, in BMVC, each block is not necessarily correlated with others.
Therefore, it is important to extract non-local information in different blocks during BMVC decoding.
Towards this end, after extensive experiments, we have found that 3D-CNN architecture is powerful to conduct this task by extracting information across non-local blocks.
However, this introduces the challenge of running time, since a 3D-CNN model usually needs $12\times \sim 25\times$ longer than its 2D-CNN counterparts ({please refer to the detailed runtime analysis in subsection~\ref{Sec:network}}).
Therefore, a trade-off between speed and quality has to be made.
To balance between decoding quality and running time, we conducted extensive experiments on 2D and 3D CNN structures and how many stages we unroll the network into.
Finally, we identified a few key observations to have fast and high quality decoding.
First, the BMVC-E2E decoder needs to incorporate at least two stages.
This is because the first stage mainly conducts an initial inpainting to the missing pixels and generates blurry results.
Additional stages are essential to retrieve the high resolution features back.
Second, 3D-CNNs are shown to be more efficient at recovering fine features from non-local blocks than 2D structures.
Especially for high compression ratio (Cr $>50$) cases, a stack of dozens of 2D-CNNs no longer provides high resolution decoded results. 
In these cases, we use two-stage 3D-CNNs in the decoder aiming for the high quality reconstruction though the speed is relatively slow,

\subsubsection{Network Structure \label{Sec:network}}
Empirically, after testing on a large number of images and videos under different scenes, we have found the following network settings in BMVC-E2E leading to a good trade-off between quality and speed:
\begin{itemize}
    \item For Cr$\le$50, the BMVC-E2E decoders have two 2D-CNN (U-Net~\cite{Unet_RFB15a} structure) stages and one 3D-CNN (RevSCI~\cite{Cheng2021_CVPR_ReverSCI} structure) stage. The U-Net stage has \{32, 64, 128, 256, 512\} filters on its down-sampling branch and \{256, 128, 64, 32, 16\} filters on its up-sampling branch. The RevSCI stage consists of ten 3D reversible blocks with 32 filter channels. After optimizing the inference phase by reducing to \texttt{float16} precision, we eventually can achieve a $70 ms$ runtime ($>14fps$) when decoding a 1080\texttimes1920 image on a server with a single GPU (Nvidia A6000).
    \item For Cr$>$50, the BMVC-E2E decoders have two 3D-CNN (RevSCI) stages, each with eleven 3D reversible blocks of 64 filter channels. Trained BMVC-E2E decoders can process a 1080\texttimes1920 image with a runtime of $252 ms$ ($\sim4fps$) per frame on the same GPU server.
The increase of decoder runtime for high compression ratio is because of the additional 3D-CNN stage in the E2E decoder, which guarantees better decoding quality though at the cost of a longer inference time.
\end{itemize}

\subsubsection{Training Details}
All BMVC-E2E decoders are trained end-to-end on the GoPRO dataset~\cite{nah2017deep} and the DAVIS HD dataset~\cite{perazzi2016benchmark} using Pytorch.
For Cr $\le 50$ cases, the BMVC-E2E decoders are trained on a GPU server with 4 Nvidia GeForce Titan XP GPUs (each one with 12GB RAM).
The high Cr BMVC-E2E decoders (Cr $>50$) are trained on GPU servers with 
2 Nvidia Tesla P40 GPUs (each one with 24GB RAM).

Note that more stages with 3D-CNN can be used and this will lead to higher decoding quality but further increase the runtime.
In addition, when the spatial size of the image changes, for instance when BMVC is being used for QHD, 4K or 8K resolutions, the mask size will be changed and the network structure might need to be adjusted as well. 

\begin{table*}[htbp!] 
 \caption{PSNR (left entry in each cell in dB) and SSIM (right entry in each cell) performance for different compression methods at a wide range of Crs on the HD image of size $1080\times 1920$. `N/A' denotes not available.} 
 \label{Table:psnr}
 \resizebox{2.0\columnwidth}{!}{
 \begin{tabular}{|c||c|c|c|c|c|c|c|c|c|c|c|} 
 \hline
 Cr ($N_b$) & 150 & 120 & 100 & 80 & 72 & 60 & 50 & 40 & 32 & 24\\
 BMVC: block size & $108\times 128$ & $108\times 160$ & $108\times 192$  & $108\times 240$ & $120\times 240$ &  $216\times 160$ & $216\times 192$ & $216\times 240$ & $270\times 240$ & $270\times 320$\\
 \hline
 \bf{BMVC-PnP}&22.89,\ 0.682&24.32,\ 0.707&25.97,\ 0.745&27.30,\ 0.781&27.96,\ 0.797&28.83,\ 0.822&29.87,\ 0.848&31.10,\ 0.877&32.23,\ 0.895&33.58,\ 0.913\\
 \bf{BMVC-E2E}&\textbf{26.59},\ 0.802&\textbf{27.38},\ 0.815&\textbf{27.80},\ 0.821&\textbf{28.81},\ 0.837&29.31,\ \textbf{0.849}&30.74,\ 0.871&29.08,\ 0.839&29.23,\ 0.843&29.98,\ 0.854&31.05,\ 0.871\\
\hline \hline
Random DS&8.95,\ 0.354&9.56,\ 0.401&10.10,\ 0.431&10.88,\ 0.460&11.33,\ 0.472&12.17,\ 0.491&13.30,\ 0.516&14.87,\ 0.555&16.61,\ 0.609&18.58,\ 0.690\\
Block CS&\textbf{26.59},\ \textbf{0.819}&26.93,\ \textbf{0.826}&27.65,\ \textbf{0.837}&27.88,\ \textbf{0.843}&28.20,\ \textbf{0.849}&28.70,\ 0.857&29.38,\ 0.865&29.83,\ 0.870&30.70,\ 0.876&31.79,\ 0.884\\
\hline
JPEG&N/A&N/A&N/A&N/A&\textbf{29.45},\ 0.837&\textbf{33.46},\ \textbf{0.900}&\textbf{35.53},\ \textbf{0.925}&\textbf{36.90},\ \textbf{0.940}&\textbf{39.62},\ \textbf{0.961}&\textbf{41.61},\ \textbf{0.971}\\
\hline
\end{tabular}}
\end{table*}

\begin{figure}
\centering
\includegraphics [width=1.0\columnwidth]{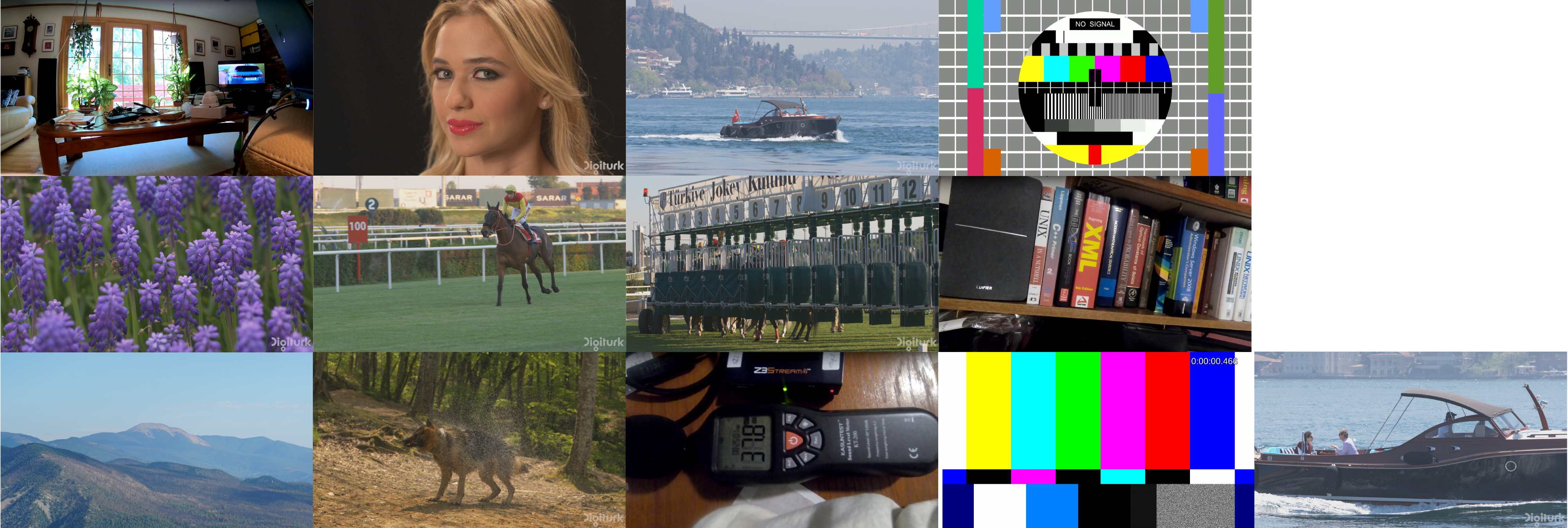}
\caption{Test dataset (\texttt{set13}) we used to evaluate the BMVC pipeline and other compression methods.}
\label{fig:set13}
\end{figure}

\begin{figure*}
\centering
\includegraphics [width=\textwidth]{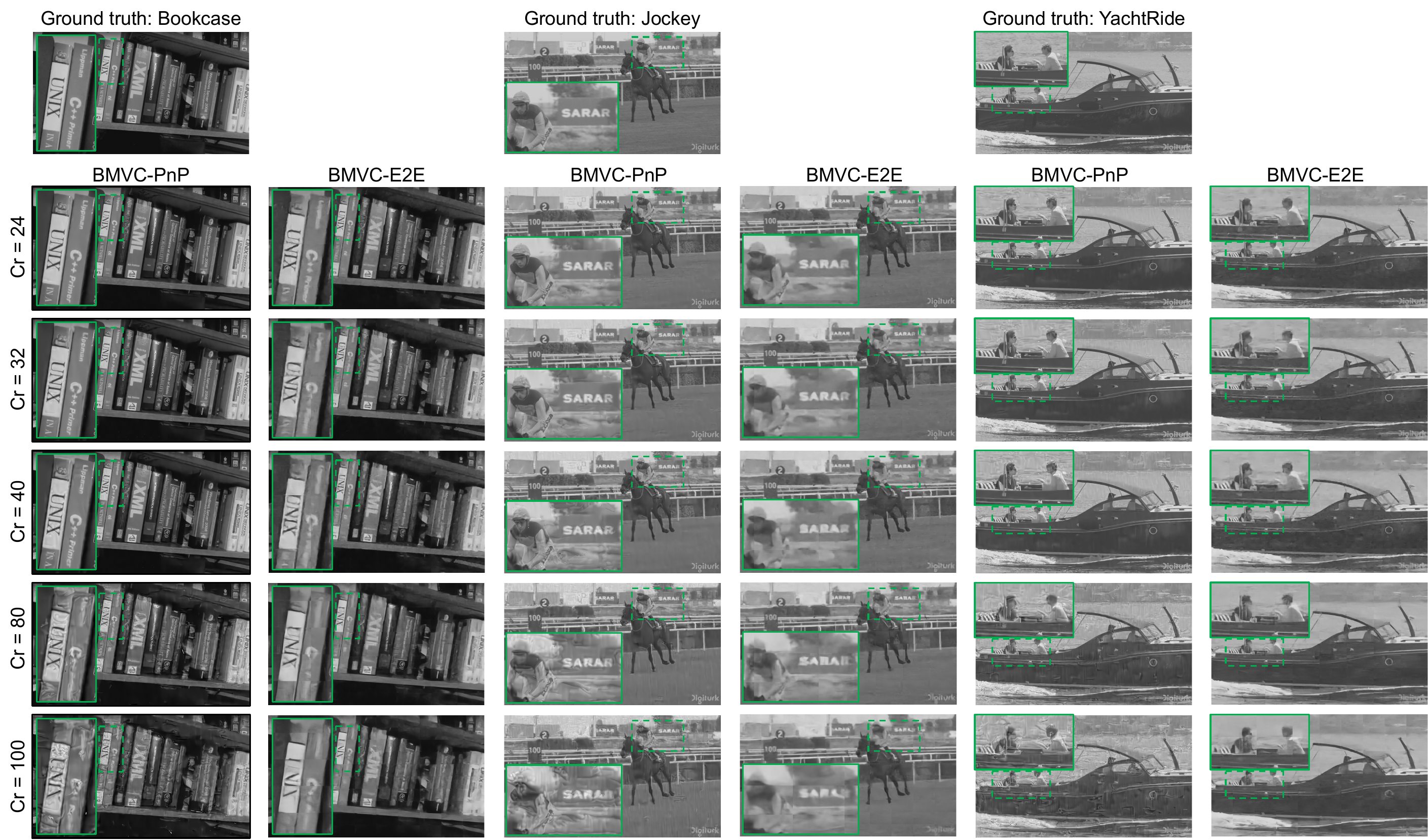}
\caption{Decoded image results at various compression ratios with the proposed BMVC-PnP and BMVC-E2E approaches. The BMVC-E2E results are consistently of good decoding quality at both low and high Crs. The BMVC-PnP decoder provides higher image quality for low Crs while produces some denoising artefacts at high Crs.}
\label{fig:BMVC_results}
\end{figure*}

\section{Evaluation Results \label{Sec:Results}}
We consider the HD video with a spatial resolution of $1080\times1920$ pixels.
The Peak Signal-to-Noise Ratio (PSNR) and Structural Similarity Index (SSIM)~\cite{Wang04imagequality} are employed as the metrics to evaluate decoded images.
For RGB videos, we first transform them to YUV videos and conduct BMVC encoding and decoding on the Y channel while only performing downsampling for U and V channels in the encoder.
In the decoder, BMVC-PnP or BMVC-E2E based algorithms are used for Y channel and bicubic interpolations are used for U and V channels respectively.
Metrics are only calculated for the restored Y channel compared with the ground truth.
We evaluate only the quality of the restored Y channel to isolate the effect of the BMVC pipeline from the interpolations of color channels.
We benchmark the BMVC pipeline along with other compression methods on static frames from the UVG dataset~\cite{mercat2020uvg} and other standard images.
Exemplar test images are shown in Fig.~\ref{fig:set13}, which covers diverse scenes.
{Examples of decoded videos using the BMVC pipeline at a wide range of Crs are presented in the Supplementary Materials. Decoded grayscale and RGB videos show the flexibility of BMVC in terms of video format.}

\begin{figure}
\centering
\includegraphics [width=1.0\columnwidth]{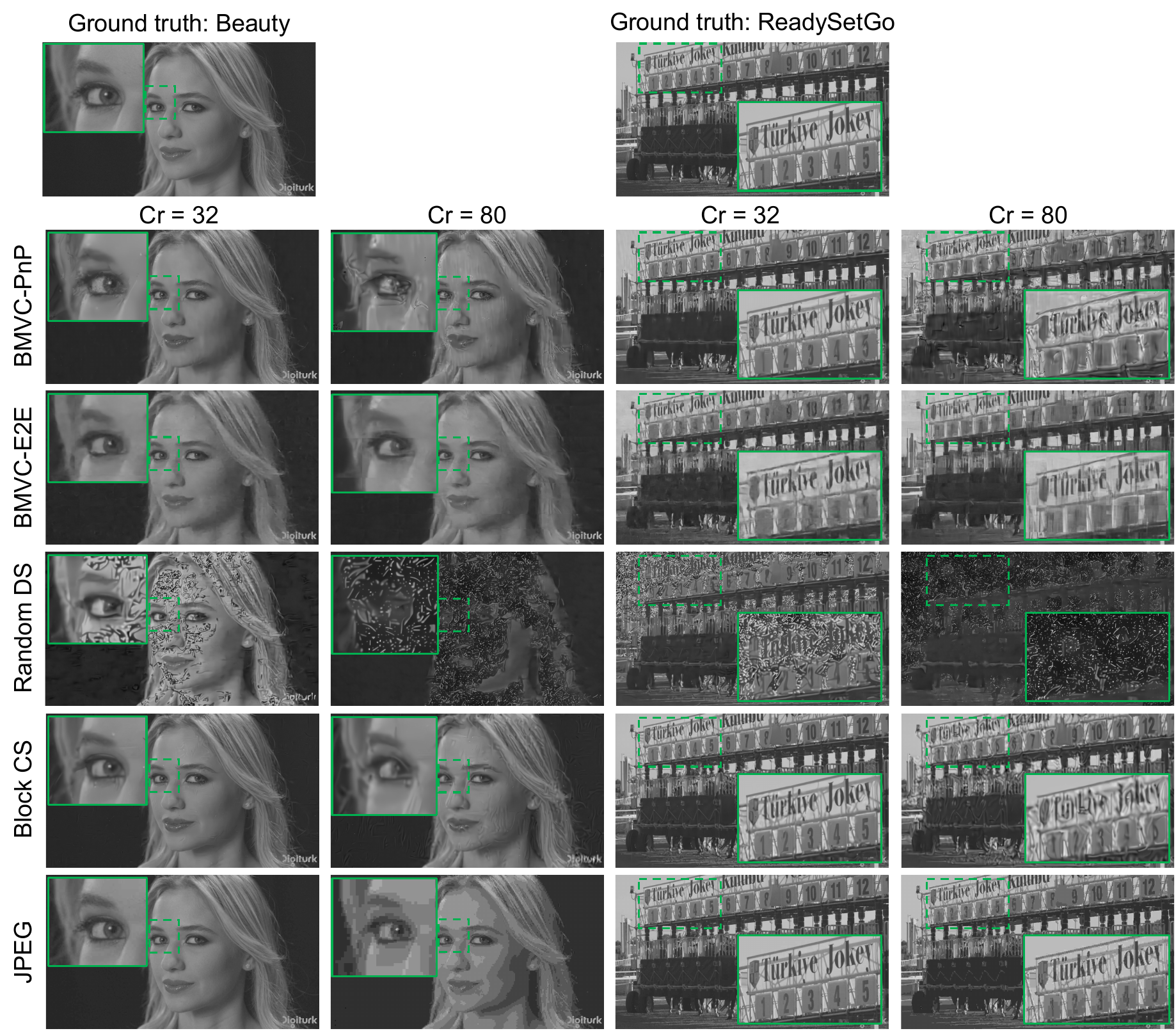}
\caption{Comparison of the BMVC pipeline with other image compression methods: Random Down-Sampling (Random DS), Block-wise Compressive Sensing (Block CS), and JPEG compression. For the Random DS and Block CS experiments, we implemented their decoders based on the PnP algorithm with FFDNet as the flexible denoiser. Results are shown with a low Cr=$32$ and a high Cr=$80$. Note that the JPEG high Cr case is Cr=$72$. The two BMVC decoders provide consistently high quality images. Random DS method fails because in principle the Random DS decoder is solving an image inpainting task with only $<3\%$ pixels available. Block CS also shows equally good decoding results but we will show later that Block CS will deteriorate after aggressive data quantization. As expected, JPEG compression gives the best image quality at Cr=$32$ since JPEG utilizes an optimal set of image basis but with the cost of increased encoding complexity. At a high Cr=$72$, JPEG starts to have some decoding artefacts due to the extremely high compression.}
\label{fig:comparison}
\end{figure}

\subsection{BMVC Decoder Evaluation at Various Compression Ratio }
Since the compression ratio of BMVC depends on the block-size, we consider the following block sizes for the HD video and the corresponding compression rations are depicted in Table~\ref{Table:psnr}.
The BMVC pipeline is tested for a wide range of Cr, ranging from 24 to 150.
The average PSNR and SSIM of the decoded images (with two different BMVC decoders) in the test set are shown in Table~\ref{Table:psnr} along with other related image compression methods, detailed in the next subsections.

Selected decoded images at representative Crs using BMVC-PnP and BMVC-E2E are presented in Fig.~\ref{fig:BMVC_results}, where we can see that for a wide range of Crs, both BMVC-PnP and BMVC-E2E decoders are consistently able to provide decent decoded images with fine details.
As expected, the resolution from BMVC starts to degrade as Cr increases.
This decreasing trend can be seen from the texts in the zoom-in panels of the `bookcase' and `jockey' examples in Fig.~\ref{fig:BMVC_results}.
In addition, we also notice the reconstruction artefacts are different for the BMVC-PnP and BMVC-E2E due to different network structures being used.

\subsection{BMVC vs. Other Compression Methods}
We further compare the BMVC pipeline with other image and video compression algorithms: Random Down-Sampling (Random DS), Block-wise Compressive Sensing (Block CS), and JPEG compression.

\begin{figure}
\centering
\includegraphics [width=1.0\columnwidth]{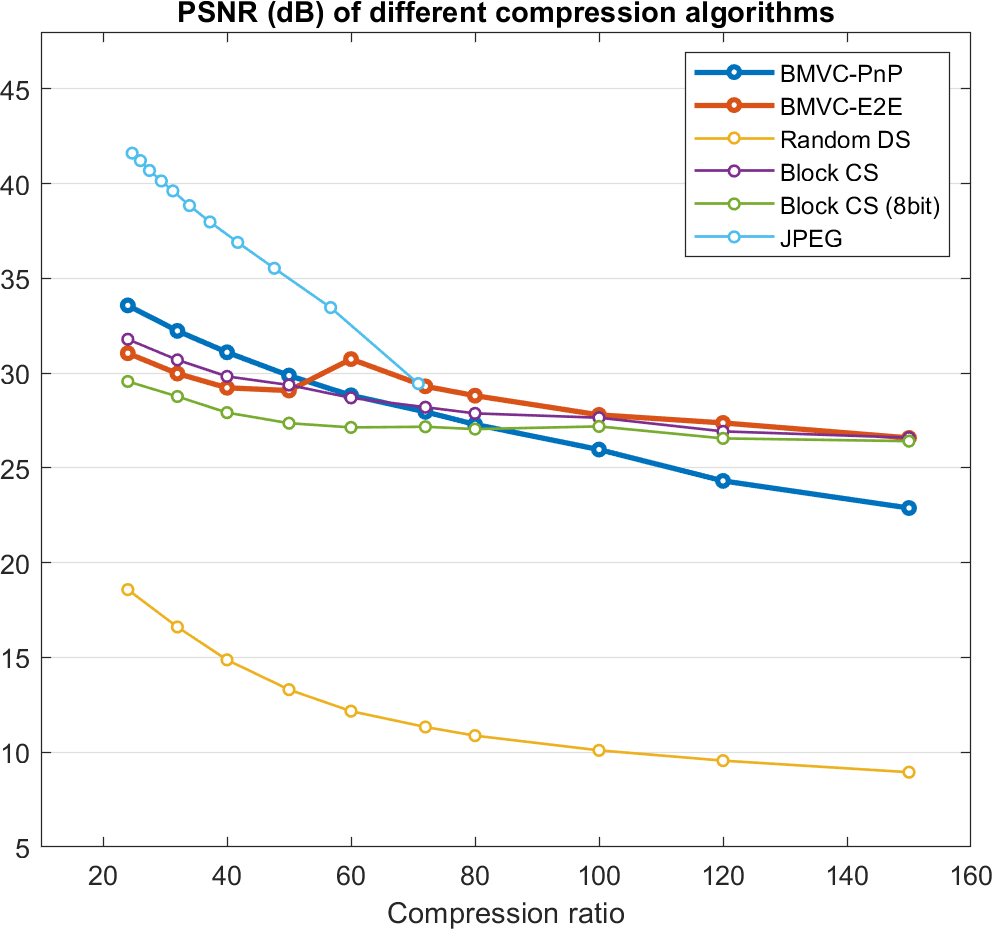}
\caption{PSNR performance of different compression methods at a wide range of compression ratios. PSNR value is computed for Y channels only. The BMVC-E2E has a PSNR increase at Cr=$60$. This is because the BMVC-E2E decoder has an additional stage of 3D CNN for all Cr$\geq 60$.}
\label{fig:psnr}
\end{figure}

\begin{itemize}
    \item {\bf Random Down-Sampling}: Since the BMVC encoding process is essentially aiming to compress the images, it is natural to first compare BMVC with Random DS which is an extreme case of compression by naively decimating some pixels in the image.
The encoder of Random DS is easily implemented by randomly sampling only $\frac{1}{Cr}$ pixels to match with the Cr of BMVC and discard all rest pixel values, which is equivalent to element-wise multiplication with a sparse binary mask.
The decoder of Random DS is implemented using the same PnP algorithm (PnP-FFDNet). 
Due to the nature of Random DS, the decoder is essentially solving an image inpainting task with extremely low counts of measured pixels ($0.6\%\sim 4\%$).
Shown in Fig.~\ref{fig:comparison}, at a relatively low Cr=$32$ case, the Random DS PnP-decoder can restore some of the spatial features but already suffers from massive reconstruction artefacts.
At a higher Cr=$80$ case, the Random DS approach entirely fails.
In terms of quantitative metrics, the PSNR performance of Random DS is also the lowest for all Crs (Fig.~\ref{fig:psnr}, Table~\ref{Table:psnr}).
We do notice the recent advanced algorithms have been developed for image inpainting (thus random DS)~\cite{ZhaTIP2021_Reconciliation}; however, these approaches are usually slow and only provide results larger than $10\%$ sampling rates. Compared with BMVC cases considered here, the compression ratios of Random DS is low.

\begin{table*}[!t] 
\centering
 \caption{Complexity of different compression methods} 
 \label{Table:complexity}
 \resizebox{\textwidth}{!}{
 \begin{tabular}{|c||c|c|c|c|l|} 
 \hline\xrowht{20pt}
 \multirow{2}{*}{Codec} & \multirow{2}{*}{Encoder Complexity} & Dynamic Range & Mask/Basis &Decoder &\multirow{2}{*}{Comment}\\
 && of  Measured Data &  Size & Complexity &\\
 \hline\xrowht{20pt}
{BMVC} & {$\frac{N}{2}{\cal O}(1)$} & {$\frac{N_b}{2}$} &{$N$} & $\begin{array}{l}
{\text{PnP:~~}} n_{iter} n \ell k^2{\cal O}(N) \\
{\text{E2E(2D):~~}} n\ell k^2{\cal O}(N) \\
{\text{E2E(3D):~~}} n \ell k^3{\cal O}(N)
\end{array}$   &$\begin{array}{l}
\text{Cr  = }N_b, N_b: \text{\# blocks }\\
\ell: \text{\# conv. layers, }n: \text{\# kernels, }k: \text{kernel size }\end{array}$\\
 \hline\xrowht{20pt}
 Random DS & ${\cal O}(1)$ & $1$ & $N$ & PnP: $n_{iter}  n \ell k^2{\cal O}(N)$ & Cr = $\frac{N}{N_s}$, $N_s$: \# sampled pixels\\
 \hline\xrowht{20pt}
 Block CS & $\frac{MN}{2}{\cal O}(1)$ & $\frac{N}{2N_b}$ & $\frac{MN}{N_b}$ & PnP: $ n_{iter} n \ell k^2{\cal O}(N)$ & Cr = $\frac{N}{M N_b}$, $M$: \# measurements per block\\
 \hline\xrowht{20pt}
  JPEG & ${\cal O}(N \log(N))$ & $1$ & $8\times 8$ & ${\cal O}(N\log(N))$& Cr depends on the quality\\
 \hline
\end{tabular}}
\end{table*}

\item {\bf Block-wise Compressive Sensing}: 
Another related CS algorithm is Block-wise Compressive Sensing (Block CS), which utilizes the same CS principle.
Unlike the traditional CS approach which gathers compressed measurements from global image pixels (such as the single-pixel camera~\cite{Duarte08SPM,Yuan2020TMM}), Block CS computes its compressed measurements only on local blocks~\cite{Zhang_2018_CVPR,Yuan18OE}; 24\texttimes24 blocks and binary sensing matrix composed of \{0,1\} are used in our experiments to be consistent with BMVC.
Note the sensing matrix is dense for Block CS (each block can be considered as a small image in the single-pixel camera) and we employ the same sensing matrix for different blocks.

For a fair comparison, we implement the Block CS decoder with the same PnP algorithm (PnP-FFDNet).
We also notice both deep learning based algorithms~\cite{You_2021_TIP_Coast} and optimization based algorithms~\cite{Zha_TIP_2021_triple} have been proposed for Block CS recently. However, comparing these algorithms is beyond the scope of this paper and later we show that BMVC outperforms Block CS on the robustness of bit quantization. 

Figure~\ref{fig:comparison} shows that the Block CS approach and its PnP-based decoder performs well in both low and high Crs.
Block CS pipeline provides visually on par results with the BMVC pipeline.
Quantitative results in Fig.~\ref{fig:psnr} and Table~\ref{Table:psnr} also show its marginal improvement over BMVC at low Crs and slightly worse scores at high Crs. 
However, a potential pitfall of Block CS is bit quantization.
Later in \ref{subsec-quantization}, we will show that BMVC is more robust to quantization bits and can provide more consistent performance after low bit quantization than Block CS.
In addition, the choice of block size  in Block CS is also important.
As the block size increases, the dynamic range of compressed data increases as well which makes the approach more sensitive to quantization.

\item {\bf JPEG Compression}: 
For decades, JPEG compression and other MPEG-based video codecs have been golden standards, providing extraordinary compression performance.
Here we test how JPEG compression performs at similar Crs of BMVC.
Unlike BMVC, Random DS, and Block CS, whose Cr can be explicitly defined, the Cr of JPEG compression can only be adjusted by its quality factor, which controls how much percentage of the most important coefficients to store after the Discrete Cosine Transform (DCT).
We adjust the quality factor to make the JPEG Cr close to BMVC.
Note that even lossless JPEG uses entropy coding to compress the data, to isolate the effect of entropy coding, we define the Cr of JPEG compression as the ratio of lossless JPEG filesize over lossy JPEG filesize.

As shown in Fig.~\ref{fig:comparison} and Table~\ref{Table:psnr}, JPEG indeed is a robust and efficient image compression algorithm in terms of both quality and compression ratio.
JPEG gives the highest PSNR performance up to Cr=$72$.
Beyond Cr=$72$, we can no longer further raise the Cr because JPEG uses an $8\times 8$ basis.
The extreme case is to store only one DCT coefficient for each block.
Despite of its superior performance, we would like to highlight that its encoding complexity is much higher than BMVC and other CS-based methods.
\end{itemize}

We summarize the decoding results of various methods in Table~\ref{Table:psnr} and Fig.~\ref{fig:psnr}.
Note that the small peak in the BMVC-E2E PSNR plot (Fig.~\ref{fig:psnr}, red) is because the BMVC-E2E decoders have an additional 3D-CNN stage for high Crs.
The SSIM metrics of the BMVC pipeline and other methods (in Table~\ref{Table:psnr}) follow a similar trend as the PSNR plot.
We can see that, at low Crs, JPEG performs best but with the price of a higher encoding cost.
The Cr of JPEG is also limited due to its $8\times 8$ basis size.
BMVC-PnP provides a higher PSNR than BMVC-E2E when Cr$\le$50 but with a longer running time.
Due to the 3D-CNN stages used in the end-to-end decoder, BMVC-E2E performs better than BMVC-PnP when Cr$>$50.
Random DS always performs the worst among these methods.
This is expected because unlike BMVC and other methods which {\em compress} the raw image, Random DS only {\em samples} a small proportion of information.
Block CS performs consistently similar to BMVC.
However, as the block size increases, the dynamic range of the compressed data increases as well.
In \ref{subsec-quantization}, we show that Block CS is sensitive to low bit quantization.
The BMVC pipeline is more consistent against different quantization bits.
Selected decoded results at various Crs and from different methods are presented in Fig.~\ref{fig:comparison}.
The visual quality and quantitative metrics (Table~\ref{Table:psnr}) both show the performance of BMVC pipeline decreases as Cr increases.
This is expected and note that a Cr above 100 is extremely challenging, not mentioning our Cr does not include entropy coding, which is part of our future work.

\subsection{Encoder and Decoder Complexity}
\label{subsec-complexity}
A key advantage of the BMVC pipeline and other CS-based methods over DCT-based image compression is its ultra low-cost encoder, which makes it perfect on resource limited mobile platforms.
Here we evaluate the computation complexity of the encoder and decoder from the compression methods discussed above, shown in Table~\ref{Table:complexity}.
Consider the original image of size $N_x$ rows and $N_y$ columns,
The total number of pixels is $N = N_x\times N_y$.
For the BMVC pipeline, let $N_b$ denote the number of blocks, which is also the Cr when there is no overlapping between blocks.
Given the modulation binary mask has equal probability being $\{0,1\}$, the expectation of the number of summations is $\frac{N_b}{2}\times \frac{N}{N_b} = \frac{N}{2}{\cal O}(1)$.
Consider the pixel values are normalized to [0,1], we can expect the encoded block has all its values bounded between $[0,\frac{N_b}{2}]$.

The corresponding BMVC decoder complexity depends on the number of iterations of the PnP approach and the CNN structure in the E2E approach.
For 2D-CNNs, the input tensor has the spatial dimension $\frac{N}{Cr}$ (rows and columns combined) and Cr channels. 
Consider a 2D convolution layer with $n$ filters of size $k\times k$.
The total amount of required multiplications to compute an output tensor with shape $\frac{N}{Cr}\times n$ is $k^2\times Cr\times \frac{N}{Cr}\times n = n k^2 N$.
The BMVC-E2E decoder with $\ell$ such 2D layers has its complexity $n\ell k^2{\cal O}(N)$.
Similarly, for 3D-CNNs, the decoder complexity is $n\ell k^3{\cal O}(N)$.
The BMVC-PnP decoder uses FFDNet with 2D layers, when running $n_{iter}$ iterations, the decoder complexity is $n_{iter} \ell n k^2{\cal O}(N)$.
In all reported FFDNet-PnP approaches (BMVC-PnP, Random DS, Block CS), we use $n_{iter} = 60$.

The Random DS approach has the lowest possible encoder complexity, which does not include any calculation but simply reading out a subset of pixel values from the detector.
Such a simple pipeline gives a complexity of ${\cal O}(1)$ and unaltered measured data bounded between $[0, 1]$.
When using the PnP framework to perform inpainting, the decoder has a complexity of $n_{iter} \ell n k^2{\cal O}(N)$.

For Block CS, let $M$ denote the number of compressed measurements recorded for each block.
Assume the sensing matrix is binary with equal probability, the number of additions in Block CS encoder is: $M\times \frac{N}{2N_b}\times N_b$, thus the encoder complexity is $\frac{MN}{2}{\cal O}(1)$.
The compressed measurement is expected to be upper-bounded by half the number of pixels in each block, which equals $\frac{N}{2N_b}$.
The PnP-decoder for Block CS uses the same framework as BMVC-PnP.
The decoder complexity is $n_{iter} \ell n k^2{\cal O}(N)$.

JPEG compression is based on DCT transform with a basis size of $8\times 8$ pixels.
The complexity of DCT transform and its inverse transform have been well studied.
Both the encoding and decoding process have a computation complexity of ${\cal O}(N\log(N))$.

\begin{figure}
\centering
\includegraphics [width=1.0\columnwidth]{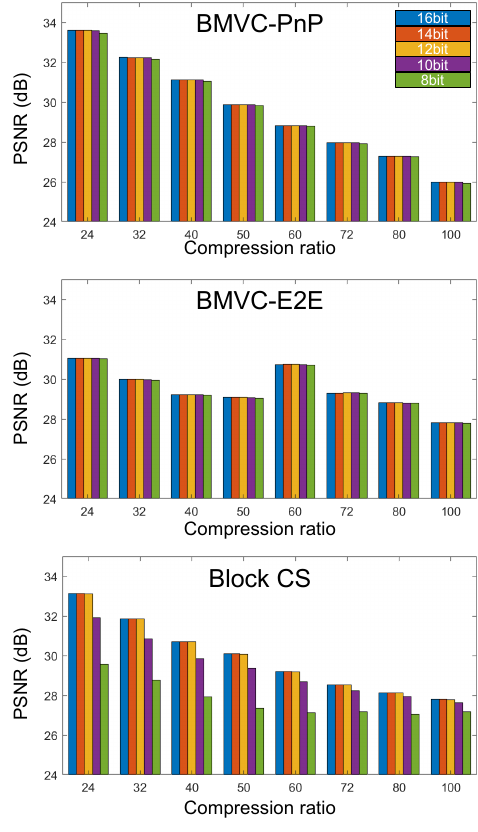}
\caption{Evaluation of robustness to quantization bits. BMVC and Block CS both show high PSNR performance when the dynamic range of the data is intact. In practice, quantization will affect the codec performance in real world video signal transmission. The bar plots indicate how the three decoders (BMVC-PnP, BMVC-E2E, and Block CS) perform under different quantization bits. BMVC decoders have consistent performance regardless of data quantization. However, {\em Block CS has noticeable decreases} of PSNR at $10bit$ and $8bit$ quantization.}
\label{fig:quantization}
\end{figure}

\subsection{Robustness to Quantization Bits}
\label{subsec-quantization}
In real world applications, the encoded data will have to be quantized before sending to the receiver.
High dynamic range data are prone to be degraded after bit quantization.
As shown in the `dynamic range' column in Table~\ref{Table:complexity}, our BMVC encoded blocks range between $0$ and $\frac{N_b}{2}$, which is upper-bounded by half the Cr.
While in Block CS, each entry of the compressed measurement may have a dynamic range up to $\frac{N}{2N_b}$, which is half the number of pixels in each block.
Under general BMVC and Block CS settings, we expect the BMVC pipeline to provide more consistent decoding performance than Block CS due to its lower dynamic range.

To test how robust the BMVC pipeline is to quantization bits, we conducted 8$\sim$16 bits quantization to the encoded/compressed blocks before sending to the two BMVC decoders and the Block CS decoder.
Decoded results are evaluated by PSNR metric.
Table~\ref{Table:quantization} and Fig.~\ref{fig:quantization} illustrates how BMVC and Block CS performs under different quantization bits.
As expected, the two BMVC decoders are robust to quantization bits.
The difference in PSNR is generally below $0.1$ dB, with only one exception for the BMVC-PnP at Cr$=24$ ($0.163$ dB degradation).
The BMVC-E2E decoder is even more consistent against quantization bits, with maximum PSNR degradation only $0.037$ dB.
Block CS has much more significant deterioration after low bit quantization.
At Cr$=24$, Block CS performance decreases by $3.59$ dB when quantized to $8bit$ data.
Similar degradation of SSIM in Block CS is observed.
Specially, Block CS at $8bit$ quantization has more reconstruction artefacts after the PnP-based decoder.
In general, Block CS requires the compressed measurements to be at least $12bit$ to secure a consistent performance.
That suggests under the same Cr, Block CS ($12bit$) will cost an extra $50\%$ bandwidth than the BMVC pipeline ($8bit$).

\begin{table*}[htbp!] 
\centering
 \caption{PSNR of BMVC-PnP, BMVC-E2E and Block CS under different quantization bits} 
 \label{Table:quantization}
 \centering
 \resizebox{\textwidth}{!}{
 \begin{tabular}{|c||c|c|c|c|c|c|c|c|c|c|c|c|} 
 \hline
  & \multicolumn{4}{c|}{BMVC-PnP (dB)} & \multicolumn{4}{c|}{BMVC-E2E (dB)} & \multicolumn{4}{c|}{Block CS (dB)}\\
 \hline
 Bit/Cr&100&80&50&24&100&80&50&24&100&80&50&24\\
 \hline
 8bit&25.921&27.274&28.829&33.451&27.785&28.796&29.060&31.023&27.190&27.049&27.361&29.563\\
 \hline
 10bit&25.981&27.305&29.882&33.592&27.805&28.813&29.082&31.056&27.634&27.957&29.361&31.931\\
 \hline
 12bit&25.981&27.301&29.889&33.611&27.803&28.818&29.087&31.059&27.800&28.123&30.090&33.132\\
 \hline
 14bit&25.982&27.303&29.892&33.613&27.802&28.819&29.088&31.060&27.813&28.132&30.099&33.149\\
 \hline
 16bit&25.982&27.304&29.879&33.576&27.803&28.818&29.088&31.060&27.814&28.133&30.102&33.151\\
 \hline
 $\Delta_{\rm PSNR} \downarrow$&0.0606&0.0315&0.0629&0.1627&0.0202&0.0230&0.0282&0.0370&\textbf{0.6238}&\textbf{1.0844}&\textbf{2.7415}&\textbf{3.5872}\\
 \hline
\end{tabular}}
\end{table*}


\subsection{Ablation Study of BMVC-E2E}
We conduct ablation study of the BMVC-E2E decoder structure to figure out the optimal design that balances well between decoder runtime and quality.
Using Cr=40 as an example, we trained multiple BMVC-E2E decoders with structures in Table~\ref{Table:ablation}.
After training, decoded results on the testing dataset is evaluated by PSNR and reported in Table~\ref{Table:ablation}.
As we employ more stages in the BMVC-E2E decoder, the decoding quality indeed improves while at the cost of longer decoding runtime.
Significant improvement is observed when adding the first stage of 3D-CNN as it is capable of efficiently gathering information from the stack of non-local blocks.
However, 3D-CNNs largely slow down the decoder runtime.
Finally, we decided to aim at $>10fps$ real-time decoding for low Crs and $>4fps$ for high Crs.
That gives us the BMVC-E2E decoder structures discussed in Sec.~\ref{sec-e2e}, which balances between decoding quality and runtime.
Notably, we also empirically found that adding more 2D CNNs does not provide on par results as 3D CNN counterparts and high Crs (Cr$>$50) benefits from using only 3D CNNs.

\begin{table}[htbp!] 
\centering
 \caption{Ablation Study of BMVC-E2E (Cr$\le 50$)} 
 \label{Table:ablation}
 \centering
 \resizebox{1.0\columnwidth}{!}{
 \begin{tabular}{|c|c|c|} 
 \hline
  Decoder structure & PSNR (dB) & Runtime (ms)\\
 \hline
 $\begin{array}{c} 
 2\times {\text{U-Net}} 
+ 2\times {\text{3D-CNN (RevSCI)}}\\
+ 1\times {\text{deeper 3D-CNN(RevSCI)}}
 \end{array}$
   & 31.720 & $\sim 255$\\
 \hline
 2$\times$ U-Net + 2$\times$ 3D-CNN (RevSCI) & 31.449 & $\sim 130$\\
 \hline
 \textbf{2$\times$ U-Net + 1$\times$ 3D-CNN (RevSCI)} & \textbf{29.228} & \textbf{$\sim 70$}\\
 \hline
 2$\times$ U-Net & 17.930 & $\sim 10$\\
 \hline
 1$\times$ U-Net & 10.308 & $\sim 5$\\
 \hline
\end{tabular}}
\end{table}

\section{Conclusions \label{Sec:Conclusion}}
To conclude, we have proposed a brand new Block Modulating Video Compression codec (BMVC) that has an ultra low-cost encoder, which is perfect for resource limited platforms such as robotics and drones.
We have also developed two BMVC decoders, based on plug-and-play optimization and end-to-end neural networks.
The BMVC-PnP decoder is an iterative algorithm that has proved convergence.
It is also flexible and robust to different compression ratios.
The BMVC-E2E decoder has an unrolled structure of the PnP approach with very few stages.
Its feed-forward nature enables real-time decoding of 1080p image sequence on a single GPU, achieving $>14fps$ for Crs under 60 and $4fps$ for higher Crs.

Unlike traditional image and video compression algorithms that embed prior knowledge in the encoding process via optimal basis selection, the BMVC pipeline takes a different design philosophy.
We only fully utilize the prior knowledge in the decoding process while keep the encoding process as simple as possible.
As a result, we are able to keep the computation complexity of the encoder as low as ${\cal O}(1)$ so that we can save power, computation and bandwidth on resource limited platforms.
On the other hand, the decoding process of BMVC is ill-posed so that it requires strong prior knowledge about natural images to reliably retrieve the desired frames.
This is achieved via either a plug-and-play framework with an image denoising CNN  or an end-to-end neural network trained on massive datasets.

One aspect we have not explored is using temporal information across frames to further compress the data stream.
It is true that state-of-the-art video codecs take advantage of the temporal redundancy of video data.
We chose to make the BMVC pipeline a frame-independent codec for three reasons: low complexity, low latency and constant bandwidth.
To compute the differential signal between frames will cost extra power consumption, which is not in line with our original purpose of developing BMVC.
Another issue raised up by processing a bunch of frames is that it will increase the latency of the end-to-end pipeline.
The third advantage of the frame-independent BMVC pipeline is its constant bit rate, which differs from the content-dependent bandwidth of MPEG-based video codecs.

Remind again that BMVC is not designed to replace the current codec standard, but to provide an alternative option for platforms under extreme power/computation constraints.
In fact, we have managed to implement the BMVC encoding pipeline on a drone platform with a simple Raspberry Pi development board.
Future work will be around building the end-to-end BMVC pipeline to achieve optimal performance (both speed and quality) and minimal latency.
With this ultra low-cost encoder design and two options of BMVC decoders, we envision the BMVC pipeline could be a revolutionary technique for video signal compression and transmission on robotic platforms.

\bibliographystyle{IEEEtran}
\bibliography{ref.bib}
\end{document}